\documentclass[12pt]{article}

\usepackage{flushend, cuted}
\usepackage{bbm}
\usepackage{amssymb}
\usepackage{float}
\usepackage{color,amsmath,amssymb, amsfonts,amstext,amsthm}

\usepackage{epsfig, graphicx, graphics}
\usepackage{subfigure}
\usepackage{longtable}
\usepackage{cite}
\usepackage[top=0.85in,left=2.75in,footskip=0.75in]{geometry}

\usepackage{changepage}
\usepackage{fixltx2e}

\usepackage[utf8x]{inputenc}

\usepackage{textcomp,marvosym}

\usepackage{cite}

\usepackage{nameref,hyperref}

\usepackage[right]{lineno}

\usepackage{microtype}
\DisableLigatures[f]{encoding = *, family = * }

\usepackage[table]{xcolor}

\usepackage{array}
\usepackage{graphicx,setspace}
\usepackage{psfrag}
\usepackage{amsfonts}
\usepackage{amssymb, amsthm}
\usepackage{mathrsfs}
\usepackage{epstopdf}
\usepackage{float}
\usepackage{lineno,hyperref}
\usepackage{color}

\makeatletter
\renewcommand{\@biblabel}[1]{\quad#1.}
\makeatother

\title{Most Probable Transition Pathways and Maximal Likely Trajectories in a Genetic Regulatory System \footnote{$\quad$$^{1}$ Center for Mathematical Sciences \& School of Mathematics and Statistics \& Hubei Key Laboratory of Engineering Modeling and Scientific Computing, Huazhong University of Science and Technology, Wuhan 430074, China.
$\quad$$^{2} $ School of Mathematics and Statistics, Henan University, Kaifeng 475000, China.
$\quad$$^{3}$ Department of Applied Mathematics, Illinois Institute of Technology, Chicago 60616, USA.
$\quad$$^{4}$School of Mathematics and Statistics, Zhengzhou University, Zhengzhou 450001, China.
$\quad$$^{a)}$ Author to whom correspondence should be addressed: huiwang2018@zzu.edu.cn; huiwheda@hust.edu.cn
}}
\author{ Xiujun Cheng$^{1}$, Hui Wang$^{4,a)}$, Xiao Wang$^{2}$, Jinqiao Duan$^{3}$, Xiaofan Li$^{3}$}

\date{}

\begin{document}
\maketitle

\begin{abstract}
We study the most probable transition pathways and maximal likely trajectories
in a genetic regulation model of   the  transcription factor activator's concentration evolution,  with Gaussian noise
and non-Gaussian stable L\'evy noise in
the synthesis reaction rate taking into account, respectively.
We compute the most probable transition pathways by the Onsager-Machlup least action principle,  and calculate the maximal likely trajectories by spatially maximizing the probability density
of the system path, i.e., the solution of the associated nonlocal Fokker-Planck equation.
We have observed  the rare most probable transition pathways in the case of Gaussian noise, for certain noise intensity, evolution time scale and system parameters.
We have especially studied  the maximal likely trajectories  starting at the low concentration metastable  state, and examined whether they evolve to or near the high concentration metastable state
(i.e., the likely transcription regime) for certain parameters,  in order to gain
 insights into the  transcription processes
 and the  tipping time for the transcription likely to occur.
 This enables us: (i)  to visualize  the progress of  concentration evolution
 (i.e., observe whether the system enters the transcription regime within a given time period);
 (ii) to predict or avoid certain transcriptions via selecting specific noise parameters in particular regions in the parameter space.  Moreover,  we have  found some peculiar or counter-intuitive phenomena in this gene model system,
    including:  (a)  A  smaller noise intensity  may trigger the transcription process, while a larger noise intensity  can not,  under the same asymmetric L\'evy noise. This phenomenon does not occur in the case of symmetric L\'evy noise;  (b) The symmetric  L\'evy motion    always  induces transition to high concentration,  but certain asymmetric    L\'evy motions  do not trigger the switch to transcription.

These findings   provide   insights for further experimental research, in order to achieve or to avoid  specific  gene transcriptions, with possible relevance for medical advances.
\end{abstract}




\section{Introduction}
\setcounter{equation}{0}

 Random fluctuations have been extensively considered in the  modeling and analysis of genetic regulatory systems \cite{Raser2005,Maheshri2007,Swain2002,Kittisopikul2010,Bressloff2014, Suel2007}.
 These fluctuations may lead to switching between  gene expression
  states  \cite{Turcotte2008,Assaf2011,Hasty2000, Hasty2000b, Liu2004, Xu2013,Zheng2016}.
  To characterize this switching behaviour, researchers have been developing   stochastic
 models \cite{Liu2004,Xu2013,Zheng2016, Augello2010,Cognata2010} by taking noises into account in deterministic differential equations.
  Noisy fluctuations are  mostly considered as Gaussian white noise in terms of
 Brownian motion \cite{Hasty2000b,Liu2004,Gui2016,Suel2006,Li2014}.
But it has been observed that the transcriptions of DNA from
genes and  translations into proteins occur in a intermittent, bursty  way \cite{Friedman2006,Lin2016,Holloway2017,Kumar2015,Dar2012}.
 This evolutionary manner  \cite{Raj2006,Golding2005,Bohrer2016,Muramoto2012,2009Dubkov,Alexander2007,Dubkov2008}
  resembles the features of  trajectories or solution paths of a stochastic differential equation with a   L\'evy motion.  In contrast to the(Gaussian) Brownian motion, L\'evy motion is a non-Gaussian   process with heavy tails and occasional jumps.

For further information about noisy gene regulation, see  \cite{Augello2010,Suel2006,Friedman2006,Lin2016,Choi2008,Tabor2008,Munsky2012,Ciuchi1993,Ciuchi1996,2018Li}
and also \cite{Gui2016,1984Horsthemke,2010Fiasconaro,2014Valenti,2003Agudov}.


In this present paper, we consider a genetic
regulatory model for the evolution of the concentration for a transcription factor activator (TF-A),
developed by Smolen et al.\cite{Smolen1998}, with the synthesis reaction rate perturbed by
stable L\'evy  fluctuations.
Liu and Jia \cite{Liu2004} investigated the effect of fluctuations arise from the
Gaussian noise in the degradation and the synthesis reaction rate of the
transcription factor activator, and found that a successive switch process occurred
with the  increase of the cross-correlation intensity between noises.
In addition, Zheng et al. \cite{Zheng2016} used the mean first
exit time  and the first escape probability
to examine the  mean time scale and the likelihood for the concentration profile
 to evolve from low concentration regime to high concentration regime
 (indicating the transcription status).

 Unlike the existing works in examining transition possibility and time scales under   noises (e.g., \cite{Liu2004, Zheng2016}), the objective of this present paper is to study the most probable transition pathways and maximal likely trajectories   for concentration states \emph{themselves} of the transcription factor activator (a protein)  as time goes on.
 This offers the following information for the gene regulation system:\\
 (i)  The concentration transition pathways and evolution trajectories from low concentration  to (or arriving near)  high concentration, indicating the evolutionary routes  to transcription.  \\
(ii)  The tipping time for the most probable transition pathways and maximal likely trajectories to pass the barrier between the
low concentration state and the high concentration state.

 To this end, we compute the most probable    transition pathways (in the case of Gaussian noise), and the maximal likely trajectories  (in the case of non-Gaussian noise)  for this stochastic gene model, in order to gain
 insights into the concentration  trajectories from the low concentration state to  the high concentration state (i.e., the likely transcription regime), and the tipping time for these trajectories passing a threshold state between low and high concentration. Especially, we try to characterize  the dependence of these dynamical behaviors on the noise parameters.

This paper is organized as follows. In Section 2, we introduce the TF-A
monomer concentration model in a gene regulation system excited by a (Gaussian) Brownian motion and (non-Gaussian)   L\'evy motion,  and propose our method on most probable transition pathways and maximal likely trajectories. In Section 3, we present the most
probable transition pathways for the gene system under Gaussian noise. Further, we consider the maximal likely trajectories for the gene system with various   non-Gaussian noise parameters, and examine which kind of  noise is more
beneficial for transcription. Finally, we summarize the above results in Section 4.
The Appendix contains basic facts about a stable L\'evy motion, and the nonlocal Fokker-Planck equation formulation for
a stochastic differential equation with a stable L\'evy motion.

\section{Model and Method}
We first introduce the TF-A monomer concentration model in a gene regulation system excited by the  (Gaussian) Brownian motion  and (non-Gaussian)  stable L\'evy motions, and then show most probable transition pathways for gene system under Gaussian noise and maximal likely trajectories.
\subsection{Model: Stochastic gene regulation}
\begin{figure}[!htp]
\begin{minipage}{0.48\linewidth}
\leftline{(a)}
\centerline{\includegraphics[height = 6cm, width = 6.5cm]{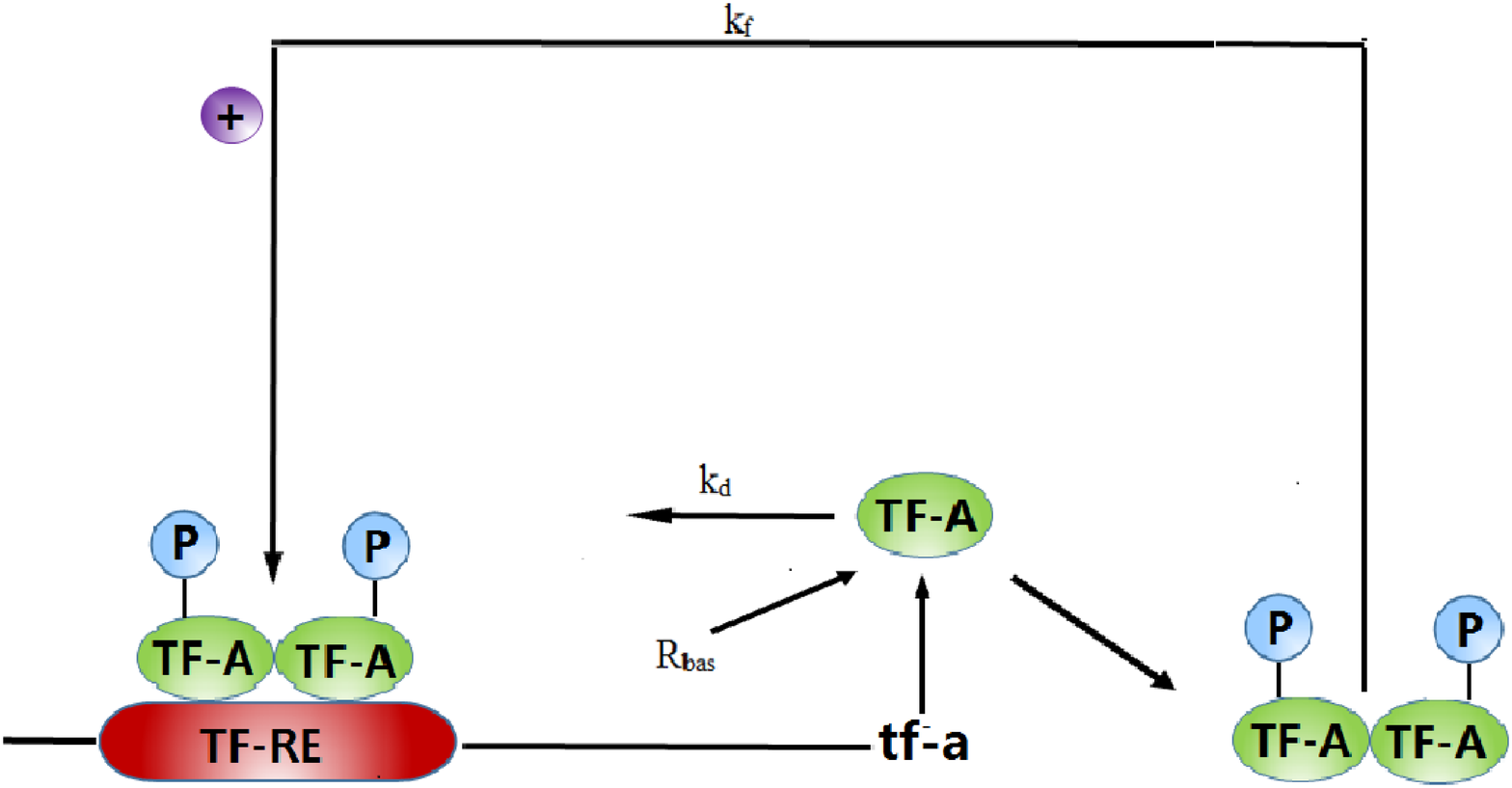}}
\end{minipage}
\hfill
\begin{minipage}{0.48\linewidth}
\leftline{(b)}
\centerline{\includegraphics[height = 6cm, width = 6.5cm]{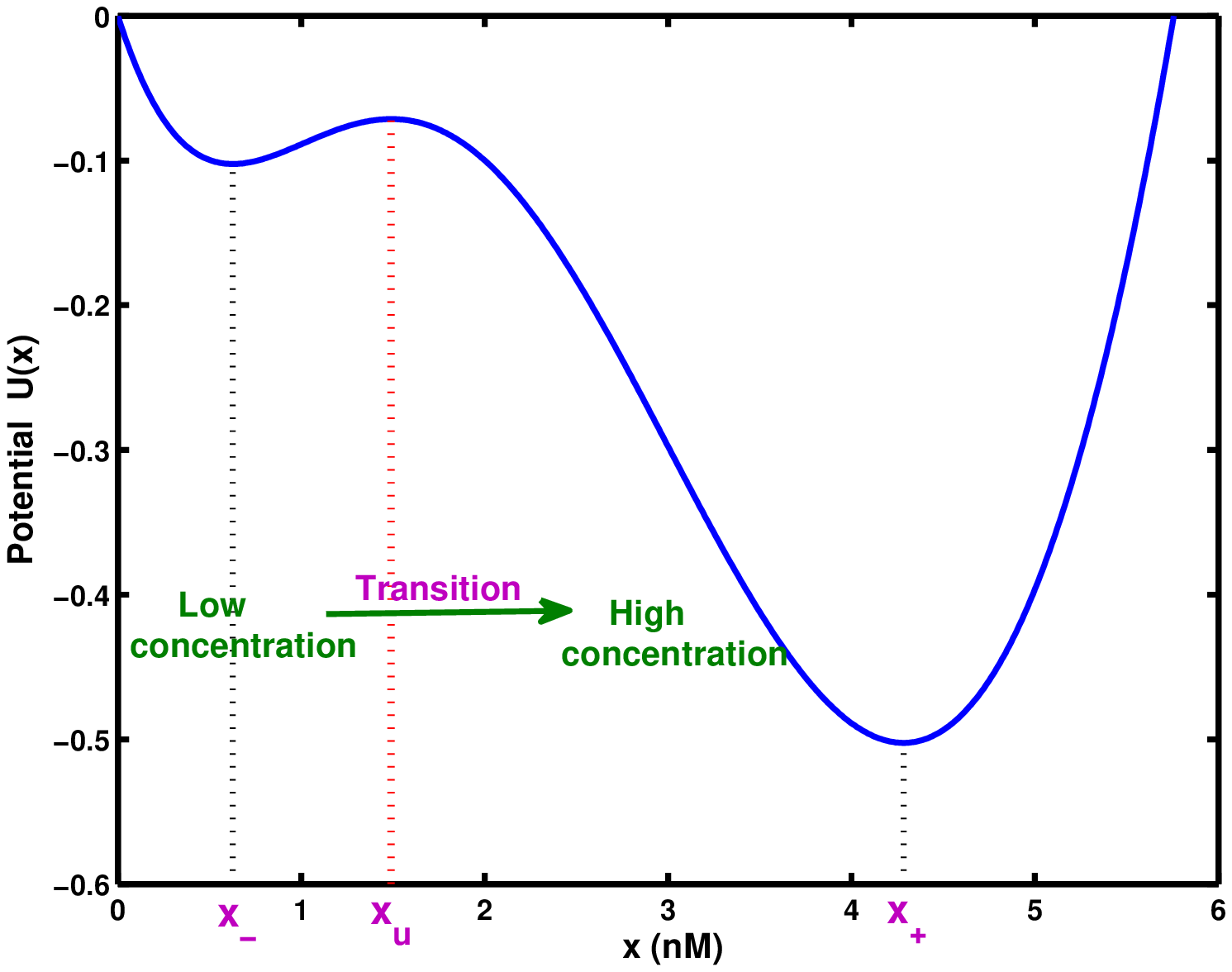}}
\end{minipage}
\caption{(a) Genetic regulatory model with a feedforward (Eq (\ref{eq:1})); (b) The bistable potential $U$ for the TF-A monomer concentration model.}
\label{Fig 1}
\end{figure}

We consider a model for the transcription factor activator (TF-A) of a
genetic regulatory system, established by Smolen et al. \cite{Smolen1998}.
A single transcriptional activator TF-A is considered as part of a pathway mediating
a cellular response to a stimulus. The TF forms a homodimer bound to responsive elements (TF-REs).
The TF-A gene incorporates one of these responsive elements,
where binding to this element of homodimers will increase TF-A transcription.
Only phosphorylated dimers can activate transcription. As  shown in Fig \ref{Fig 1}(a),
the phosphorylated (P) transcription factor activator (TF-A) activates transcription with a maximal rate $k_f$, with $k_d$ and $R_{bas}$   the degradation and synthesis rate of the TF-A monomer, respectively. The dissociation concentration of the TF-A dimer from TF-REs is denoted by $K_d$. Then the evolution of the TF-A concentration $x$ obeys the following differential equation \cite{Smolen1998}:

\begin{equation} \label{eq:1}
  \dot{x} = \frac{k_f x^2}{x^2 + K_d} - k_d x + R_{bas},
\end{equation}

 The system (\ref{eq:1}) can be rewritten as $\dot{x} = f(x)$, where $f(x) = \frac{k_f x^2}{x^2 + K_d} - k_d x + R_{bas}$.  The  vector field $f(x)$ is also equal to $-U'(x)$,  with the potential function $U(x) =k_f\sqrt{K_d}\ \arctan\frac{x}{\sqrt{K_d}} + \frac{k_d}{2}x^2 - (R_{bas} + k_f)x $.
 This system has two stable states and one  unstable state, when the parameters satisfy
 the condition: $[-(\frac{k_f + R_{bas}}{3k_d})^3 + \frac{K_d(k_f + R_{bas})}{6k_d} - \frac{K_d R_{bas}}{2k_d}]^2 + [\frac{K_d}{3} - (\frac{k_f + R_{bas}}{3k_d})^2]^3 < 0$. As in \cite{Liu2004}, we select suitable parameters:  $k_f = 6\;{\rm min}^{-1}, ~K_d = 10, ~k_d = 1 \;{\rm min}^{-1}$, and $R_{bas} = 0.4 \;{\rm min^{-1}}$. Then two stable states are $x_- \approx 0.62685 \;{\rm nM}$ and $x_+ \approx 4.28343 \;{\rm nM}$ and the unstable state (a saddle point) is $x_u \approx 1.48971 \;{\rm nM}$. See Fig \ref{Fig 1}(b).


The dynamical system (\ref{eq:1}) is a deterministic model. Some experiments indicated that basal synthesis rate $R_{bas}$ is influenced by random fluctuations arising from the biochemical reactions, the concentrations of other proteins, and gene mutations \cite{Raj2006, Raj2009, Liu2004}.
So as in \cite{Liu2004}, we consider the system \eqref{eq:1} with Gaussian noise in terms of Brownian motion
\begin{equation}\label{eq:2.01}
  \dot{X}_t = \frac{k_f X_t^2}{X_t^2 + K_d} - k_d X_t + (R_{bas} + \epsilon \dot{B_t}),   \qquad  X_0 = x_0,
\end{equation}

However, there exists powerful evidences indicating that gene expression with small diffusion and large bursting resembles the composition of systems with L\'evy noise  \cite{Raj2006,Golding2005,Bohrer2016, Muramoto2012}. Furthermore, some stochastic models of gene expression illustrate that transcriptional bursts display a heavily skewed distribution \cite{Raj2006}.
Zheng et al. \cite{Zheng2016} model
random behaviour of the synthesis rate $R_{bas}$ on the genetic regulatory system (\ref{eq:1}) by a symmetric stable L\'evy noise. But this symmetry is quite an idealized, special  situation and does not have the property of the skewness. Hence, we take an asymmetric stable L\'evy noise with skewness characteristic as a random perturbation of the synthesis rate $R_{bas}$. Then the model (\ref{eq:1}) becomes  the following scalar stochastic gene regulation model with additive stable noise:
\begin{equation}\label{eq:2.1}
  \dot{X}_t = \frac{k_f X_t^2}{X_t^2 + K_d} - k_d X_t + (R_{bas} +  \dot{\tilde{L_t}}^{\alpha,\beta}),   \qquad  X_0 = x_0,
\end{equation}
where the scalar stable L\'evy motion $\tilde{L_t}^{\alpha,\beta}$, with non-Gaussianity index $\alpha (0 < \alpha < 2)$, skewness index $\beta (-1\leq \beta \leq 1)$, scaling index $\sigma (\sigma >0)$ and shift index zero, is recalled in \nameref{S1_Appendix} at the end of this paper. We denote $\epsilon \triangleq  \sigma^\alpha$ the noise intensity. It is worth noting that we are concerned with the internal noise \cite{Liu2009,Thattai2001}, which can be realized as additive noise. In stochastic dynamical systems, it is customary to denote a system variable by a capital letter with time as subscript ($X_t$). A stable L\'evy motion is asymmetric when $\beta \neq 0$ and symmetric when $\beta = 0$.
A solution orbit (also called a solution trajectory) $X_t$ has occasional (up to countable) jumps for almost all samples (i.e.,  realizations), except in the case with Brownian motion $B_t$ for which almost all the solution trajectories are continuous in time \cite{Applebaum2009}.


  Without noise,  the low concentration stable state  $x_-$ and high concentration stable state  $x_+$  are resilient (see Fig
  \ref{Fig 1}(b)): the TF-A   concentration states will locally be attracted  $x_-$ or  $x_+$, as time increases for the deterministic system (\ref{eq:1}).   It is known that stochastic fluctuations     may induce switches between two stable states   \cite{Smolen1998,Liu2004}. The TF-A concentration,    starting near the low concentration  state  $x_-$ in $D = (0,x_u)$, arrives at a high concentration state  (near $x_+$) by passing through the unstable saddle  state  $x_u$ (see Fig \ref{Fig 1}(b)).  The \emph{threshold time instant} when the system passes the    unstable  saddle state   $x_u  \approx 1.48971$  is called the `tipping time' for a specific solution trajectory.

We now examine these system trajectories  or orbits  for the gene regulation model: \\
(i) How does the system evolve  from low concentration (near $x_{-}$) to high concentration
(near $x_{+}$), indicating the system   in  a transcription regime?  \\
(ii) What is the tipping time  while evolving from the low concentration state to the high concentration state?

\subsection{Method}
We first discuss the\emph{ most probable transition pathways} for  the genetic system with (Gaussian) Brownian motion, and then consider the \emph{maximal likely trajectories} for this system with (non-Gaussian) L\'evy motion.

\subsubsection{Most probable transition pathway}

For  the stochastic gene system \eqref{eq:2.01} with Brownian motion,  we   obtain the most probable transition pathway \cite{1978Bach}  from the  low concentration metastable state  $x_{-}$  to the high concentration
metastable state $x_{+}$,  by minimizing the   Onsager-Machlup action functional. This approach is not yet available for stochastic systems with L\'evy motions.
The most probable transition pathway       $z_m(t)$  can be obtained by the least action principle
\begin{align}
 \delta \int_{0}^{T}OM(\dot{z}, z)dt  = 0,
\end{align}
where the Onsager-Machlup function $OM(\dot{z}, z)$ is defined by
\begin{align}\label{eq:5.1}
OM(\dot{z}, z) = \left(\frac{f(z) - \dot{z}}{\epsilon}\right)^2 + f'(z),
\end{align}
and the integral $\int_{0}^{T}OM(\dot{z}, z)dt $ is called the  Onsager-Machlup action functional.
It needs to satisfy  the Euler-Lagrange equation
\begin{align}
\frac{d}{dt} \frac{\partial OM(\dot{z}, z)}{\partial \dot{z}} = \frac{\partial OM(\dot{z}, z)} {\partial z}
\end{align}
as the differential equation for $z_m(t)$. Hence,  the most probable transition pathway $z_m(t)$ satisfies  the following equation
\begin{align}
& \ddot{z}_m(t) = \frac{\epsilon^2}{2}f''(z_m) + f'(z_m)f(z_m),~~~t \in (0, T),\label{mp01}\\
& z_m(0) = x_-,~~ z_m(T) = x_+.  \label{mp02}
\end{align}
 Note that the Onsager-Machlup function given in \eqref{eq:5.1} is convex in the variable $\dot{z}$. Thus, by
Theorem 10 in Section 8.2.5 of [9, Page 481], if   there exists a $C^2$ solution to the Euler-Lagrange equation,  then it is indeed  a (local) minimizer.  We adopt a shooting method in \cite{1976Keller} to  numerically  solve two-point boundary value problem \eqref{mp01}-\eqref{mp02}.  As this boundary value problem may not have a solution, the most probable transition pathway may not exist.

\subsubsection{Maximal likely trajectory}
The Onsager-Machlup action functional approach, in the previous subsection, for the most probable transition pathway is not yet available for stochastic systems with L\'evy motions.   Instead, we consider  the maximal likely trajectory for the stochastic system \eqref{eq:2.1} with non-Gaussian  L\'evy motion,  starting at an initial state $x_0$  (but we will take it as the low concentration metastable state  $x_{-}$).  This is reminiscent of  studying a deterministic dynamical system by examining the evolution of  its trajectory  starting from an initial state.       Each sample  solution path starting at this initial state is a possible outcome of the solution path  $X_t$. What is the maximal likely trajectory of $X_t$? In order to answer the question, we need to decide on the maximal likely position $x_m(t)$  of the system (starting at  the initial point $x_0$)  at every given future time $t$, but  this  is the maximizer for the probability density function $p(x, t) \triangleq p(x, t; x_0, 0) $ of  solution $X_t$.   We first numerically solve  the  nonlocal Fokker-Planck equation  (\ref{eq:2.2}), and  then we find the maximal likely position $x_m(t)$ as the maximizer of $p(x, t)$ at every given time $t$.
 The probability density function $p(x,t)$ is a surface in the $(x,t,p)-$space. At a given time instant $t$, the maximizer $x_m(t)$ for $p(x,t)$ indicates the  maximal likely location of this orbit at time $t$. The trajectory (or orbit) traced out by $x_m(t)$ is called the maximal likely trajectory  starting at $x_0$. Thus, $x_m(t)$ follows the top ridge or plateau of the surface in the $(x,t,p)-$space as time goes on.
 Starting at every initial point, we may thus compute its maximal likely trajectory for the evolution of concentration for the transcription factor activator (and thus we occasionally call it the maximal likely evolution trajectory).
 The maximal likely trajectories \cite{Duan2015, Zhuan} are also called `paths of mode' in climate dynamics and data assimilation \cite{Miller1999, Gao2016b}.  They may have jumps,   as the  solution sample paths of the stochastic system \eqref{eq:2.1} have jumps due to L\'evy motion.

\medskip

As in \cite{Wang2018}, \emph{the maximal likely equilibrium state} is defined as a state which either attracts or repels all nearby
orbits. When it attracts all nearby orbits, it is called a maximal likely stable equilibrium state, while if it repels all nearby orbits, it is called a maximal likely unstable equilibrium state. Maximal likely equilibrium states depend on noise parameters $\alpha, \beta$, noise intensity $\epsilon$ as well as the genetic system parameters.

\begin{figure}[!ht]
\vfill
\begin{minipage}{0.48\linewidth}
\leftline{(c)}
\centerline{\includegraphics[height = 6cm, width = 6.5cm]{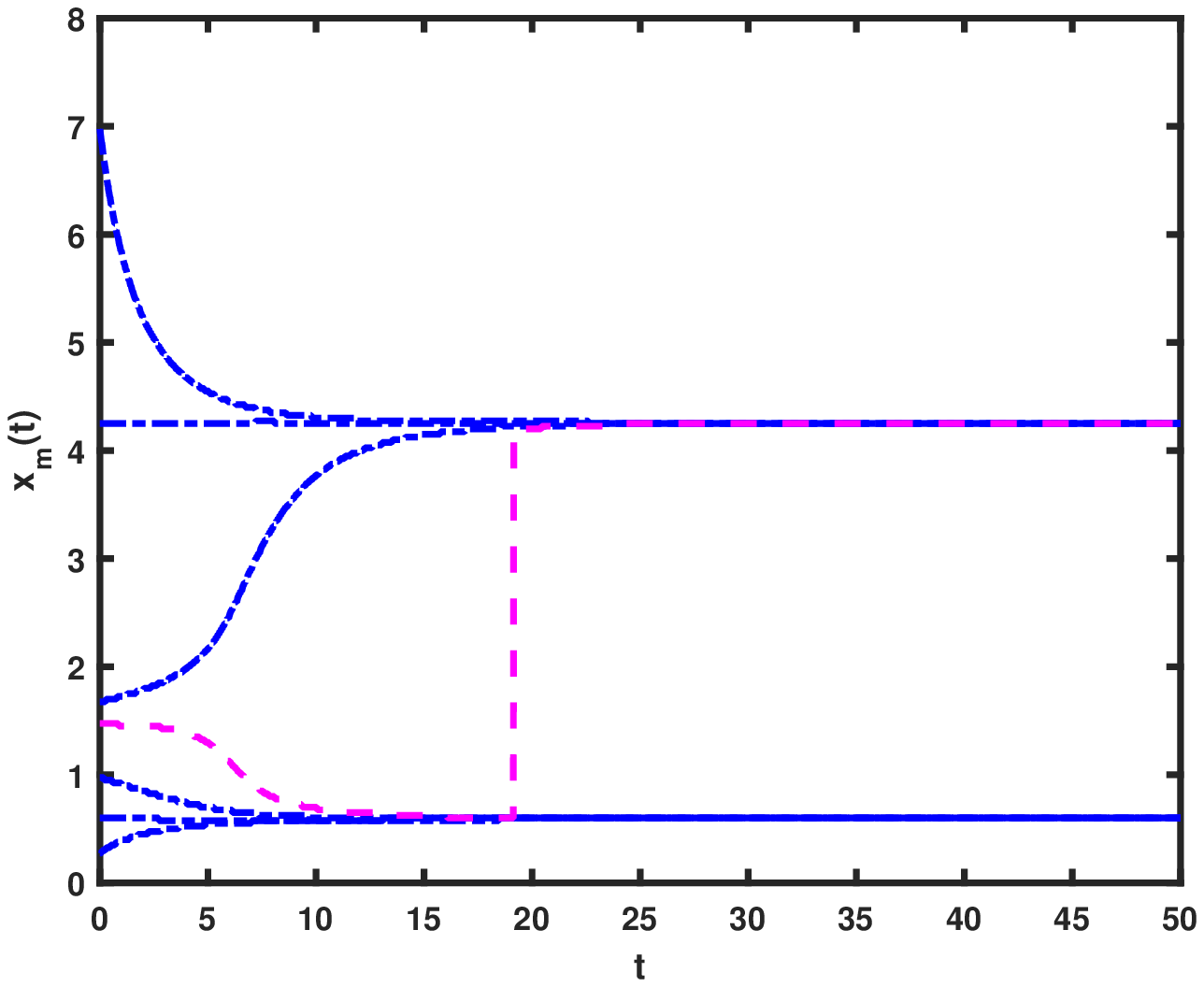}}
\end{minipage}
\hfill
\begin{minipage}{0.48\linewidth}
\leftline{(d)}
\centerline{\includegraphics[height = 6cm, width = 6.5cm]{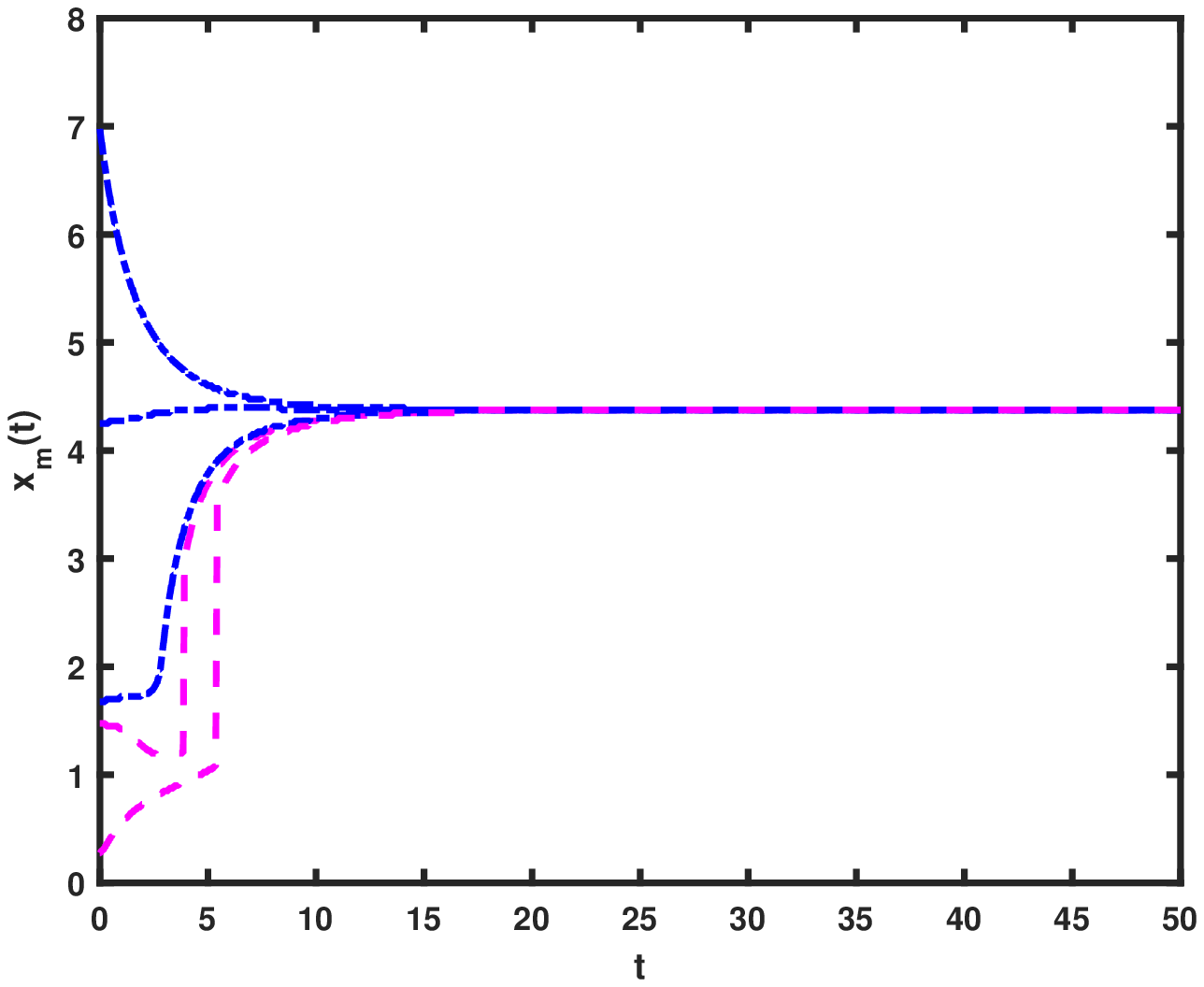}}
\end{minipage}
\caption{\textbf{(Online color) Maximal likely evolution trajectories of stochastic system (\ref{eq:2.1}) starting at various initial concentration $x_0$.} (a) With  L\'evy motion:  $\alpha = 1.5, \beta = 0, \epsilon = 0.02$. (b) With   L\'evy motion: $\alpha = 1.5, \beta = 0, \epsilon = 0.4$.}
\label{Fig5}
\end{figure}

Fig \ref{Fig5} also shows one or two maximal likely equilibrium states.  We observe that the maximal likely equilibrium state in high concentration is between $4$ and $5$, depending on $\alpha, \beta, \epsilon$, and it differs from the deterministic stable state $x_+ \approx 4.28343$ due to the effect of noise.

\medskip

In the following,  we are especially interested in the \emph{maximal likely trajectories}   starting at the low concentration stable state $x_{-} \approx 0.62685$ and approaching (or arriving in a small neighborhood of)  the maximal likely equilibrium stable state in the high concentration regime (more likely for transcription). Fig \ref{Fig5} also shows the maximal likely evolution trajectories for certain parameters.  This definition of maximal likely evolution trajectories is based on maximizing the solution's  probability density at every time instant. We use a similar efficient numerical finite difference method developed  by  us in Gao et al. \cite{Gao2016} to simulate the nonlocal Fokker-Planck  equation (\ref{eq:2.2}). This applies to stochastic systems with finite as well as small noise intensity.

\bigskip

In summary, there are significant differences between most probable transition pathway and maximal likely trajectory. Firstly, the most probable transition pathway is a continuous trajectory from one metastable state to another metastable state. The maximal likely trajectory  is a trajectory starting from one initial state. Secondly, the former is obtained via numerically solving the two-point boundary value problem (i.e., shooting method). The latter is calculated by solving an initial value problem. Thirdly, the former can be understood as the probability maximizer that sample solution paths lie  within a   tube with $z_m(t)$ as the center. The latter is determined by maximizing the probability density function (i.e., the solution of the Fokker-Planck equation) at every   time instant.

\section{Results}

In the following section, we compute the most probable transition pathways and maximal likely evolution trajectories, in order to analyze how the  TF-A concentration  evolves  from the low concentration state to the high concentration state. The most probable transition pathways $z_m(t)$ starting from $x_- \approx 0.62685$ and ending at $x_+ \approx 4.28343$ and maximal likely trajectories $x_m(t)$ starting at $x_- \approx 0.62685$,  are deterministic estimators as time goes on.   The tipping time is the time needed (counting from the start)  for this most probable transition pathways $z_m(t)$ and maximal likely evolution trajectories $x_m(t)$ to pass through the saddle point $x_u$. The tipping time for the most probable transition pathways and maximal likely evolution trajectories provides the threshold time instant  at which the system enters the high concentration regime.

\subsection{Gene regulation under Gaussian Brownian motion:  Most probable transition pathway }

\begin{figure}[!ht]
  \centering
  \includegraphics[width=8cm]{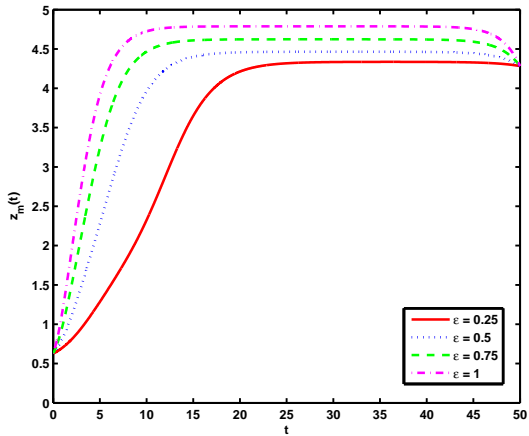}\\
  \caption{\bf{Most probable transition pathways $z_m(t)$ starting  at the low concentration metastable state $x_- \approx 0.62685$ and ending at the high concentration metastable state  $x_+ \approx 4.28343$ under Gaussian noise with different noise intensities:  $\epsilon = 0.25,~0.5,~0.75,~1$.}}\label{Fig3.0}
\end{figure}

For gene regulation system \eqref{eq:2.01}, we now examine the   most probable transition pathway   $z_m(t)$ starting at the low concentration metastable state $x_- \approx 0.62685$ and ending at the high concentration metastable state $x_+ \approx 4.28343$. As seen in Fig \ref{Fig3.0}, as time increases, for four different noise intensities $\epsilon = 0.25,~0.5,~0.75,~1$, the most probable concentration increases to the high concentration quickly, and remains a nearly constant level (in transcription regime), then decreases to the high stable concentration state $x_+$ at $T = 50$. Moreover, we clearly observe that the most probable concentration passes the `barrier'  $x_u  \approx 1.48971$ faster for larger $\epsilon$. That is, the tipping time is shorter for larger noise intensity. The existence of such rare most probable transition pathways depends crucially on the system running time  $T$, system parameters, and the  noise intensity.


\subsection{Gene regulation under non-Gaussian L\'evy motion: Maximal likely trajectory}

\begin{figure}[!htp]
\begin{minipage}{0.48\linewidth}
\leftline{(a)}
\centerline{\includegraphics[height = 6cm, width = 6.5cm]{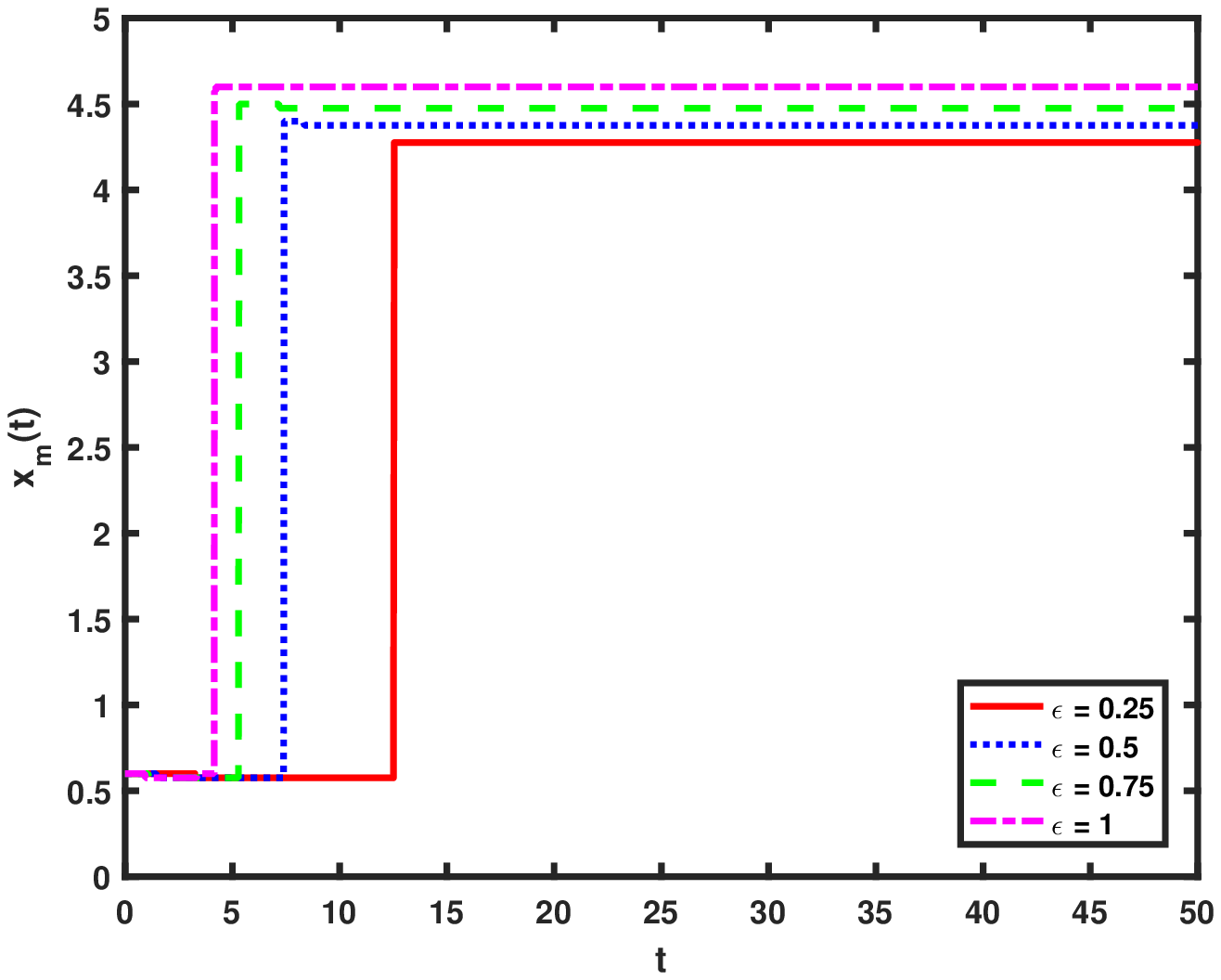}}
\end{minipage}
\hfill
\begin{minipage}{0.48\linewidth}
\leftline{(b)}
\centerline{\includegraphics[height = 6cm, width = 6.5cm]{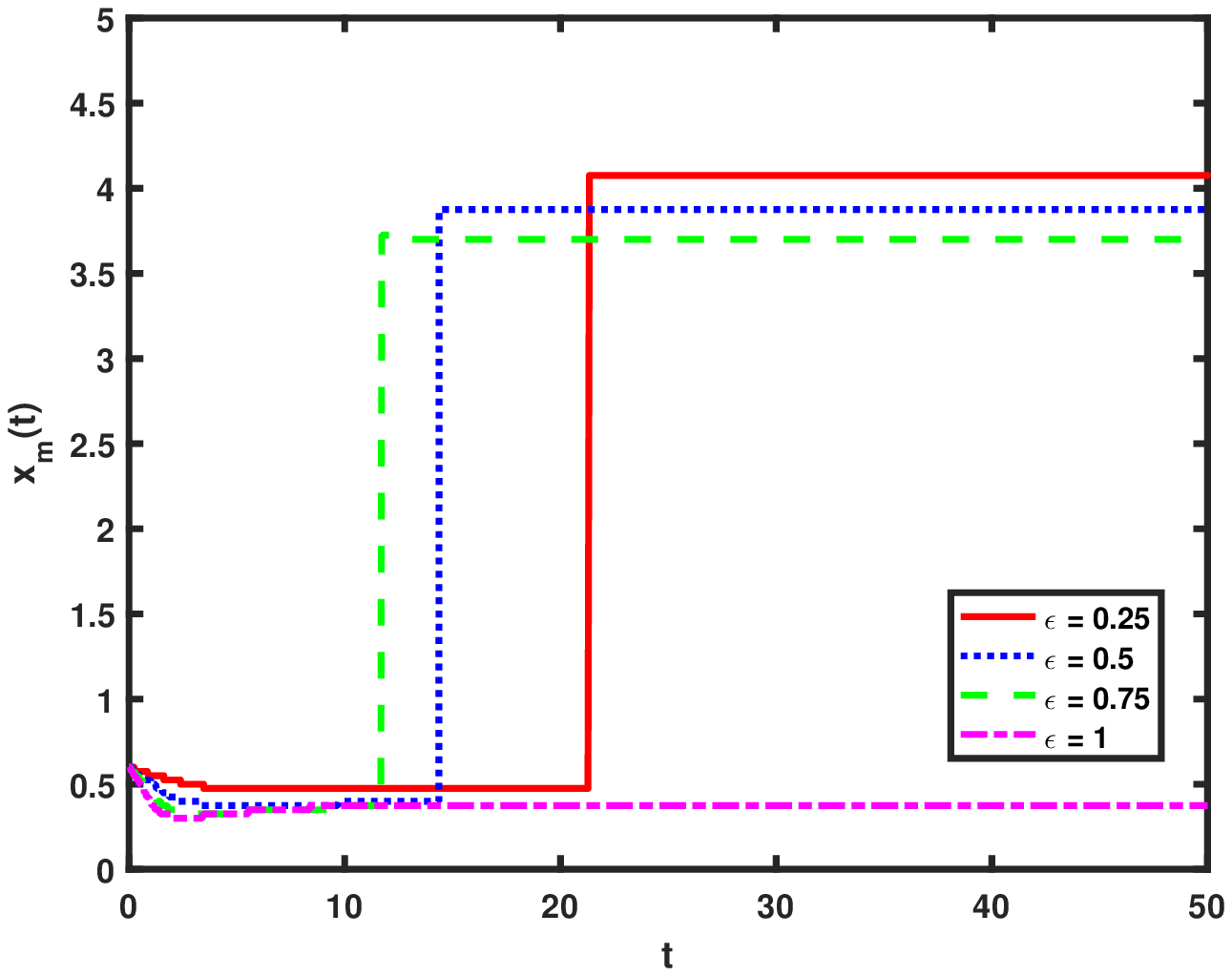}}
\end{minipage}
\caption{\textbf{(Online color) Maximal likely evolution trajectories under symmetric ($\beta = 0$) and asymmetric ($\beta \neq 0$) L\'evy noise for various parameters.}  (a) Dependence on  $\epsilon $: $\alpha = 0.5, ~\beta = 0$. (b) Dependence on   $\epsilon $: $\alpha = 0.5, ~\beta = -0.5$.}
\label{Fig7}
\end{figure}

For gene regulation system \eqref{eq:2.1}with L\'evy motion,   the Onsager-Machlup action functional approach is not yet available.  Therefore, we here  focus on   its  maximal likely trajectories.
We now investigate the influence of symmetric (see Fig \ref{Fig7}(a)) and asymmetric (see Fig \ref{Fig7}(b)) L\'evy noise on maximal likely evolution trajectories. In Fig \ref{Fig7}(a), as time increases, for four different noise intensities $\epsilon = 0.25,~0.5,~0.75,~1$, the maximal likely concentration decreases a bit at the beginning, increases to the high concentration quickly, and finally remains a nearly constant level (in transcription regime). These dynamical behaviors, in the symmetric noise case,  may appear intuitively correct, but this is not true  in the asymmetric noise case as we now discuss.

In Fig \ref{Fig7}(b), we notice that the  maximal likely concentration goes through the saddle state $x_u$ and arrives at the high concentration state for smaller noise intensities  $\epsilon = 0.25,~0.5,~0.75$, but counter-intuitively decrease to a nearly constant low concentration for larger noise intensity $\epsilon = 1$. In this work, we take time $t = 50$ as the tipping time if the  maximal likely concentration does not pass through the saddle point $x_u$ by  time $t = 50$.
This suggests that the asymmetric L\'evy noise with $\alpha = 0.5, ~\beta = -0.5$,   and  $\epsilon = 1,~x_0 = 0.62685$, does not induce the switch mechanism for transcription.

\begin{figure}[htp]
\begin{minipage}{0.48\linewidth}
\leftline{(a)}
\centerline{\includegraphics[height = 6cm, width = 6.5cm]{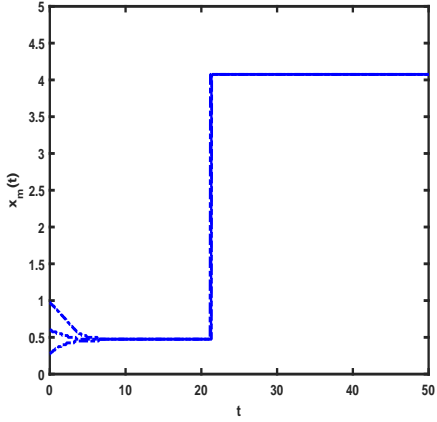}}
\end{minipage}
\hfill
\begin{minipage}{0.48\linewidth}
\leftline{(b)}
\centerline{\includegraphics[height = 6cm, width = 6.5cm]{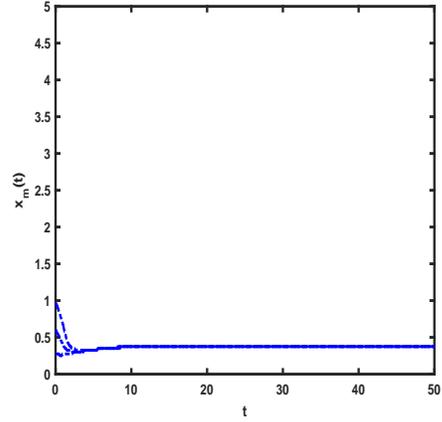}}
\end{minipage}
\caption{\textbf{(Online color) Maximal likely evolution trajectories starting at various initial concentration $x_0$.} (a) With  L\'evy motion:  $\alpha = 0.5, \beta = -0.5, \epsilon = 0.25$. (b) With   L\'evy motion: $\alpha = 0.5, \beta = -0.5, \epsilon = 1$.}
\label{Fig8}
\end{figure}

The reason for no possible transcription may be explained by the stability of two deterministic stable states $x_-,~x_+$   in system (\ref{eq:1}), under the influence of noise. As shown in Fig \ref{Fig8}(a),  the state near $x_-$  is attracted to the  maximal likely equilibrium stable state in high concentration domain and thus we have possible transcription for the smaller noise intensity $\epsilon =0.25$.    But in Fig \ref{Fig8}(b),  the state near $x_-$  is attracted to the  maximal likely equilibrium stable state  in the low concentration state,  and this offers an explanation for the counter-intuitive phenomenon of no possible transcription  for the larger noise intensity $\epsilon =1$.

In summary, by examining the  maximal likely  trajectory starting from the low concentration metastable state $x_-$ , we have thus found that   the symmetric ($\beta=0$)  L\'evy noise  appears to induce transition to transcription,  while asymmetric  ($\beta \neq 0$)  L\'evy noise with certain parameters do not trigger the switch to transcription. Furthermore, we have explained the reason for no apparent transcription, due to the system stability change  for some asymmetric stable L\'evy noise. In the next  subsection, we will further show effects of stable L\'evy noise by computing the tipping time and the concentration values for maximal likely evolution trajectories at the end of computation ($t=50$).

\subsection{Maximal likely transcription and tipping time}
\begin{figure}[!htp]
\begin{minipage}{0.48\linewidth}
\leftline{(a)}
\centerline{\includegraphics[height = 6cm, width = 6.5cm]{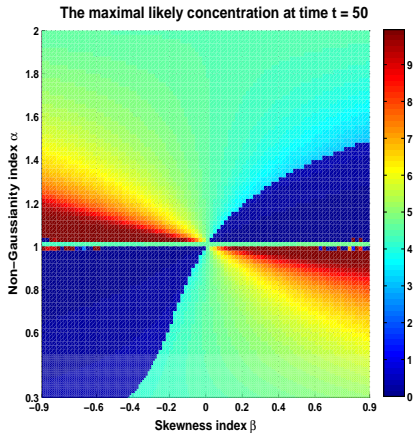}}
\end{minipage}
\hfill
\begin{minipage}{0.48\linewidth}
\leftline{(b)}
\centerline{\includegraphics[height = 6cm, width = 6.5cm]{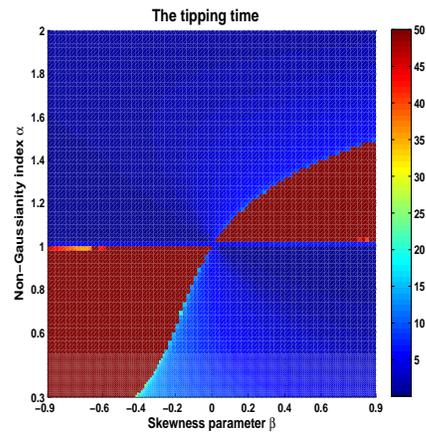}}
\end{minipage}
\caption{\textbf{(Online color) The effects of noise parameters $\alpha, \beta$ on the maximal likely concentration and the tipping time.} (a) The maximal likely concentration at $t = 50$ with initial concentration $x_0=0.62685,~\epsilon = 1$. (b) The tipping time for the maximal likely evolution trajectories with initial concentration $x_0=0.62685,~\epsilon = 1$.}
\label{Fig10}
\end{figure}

In this subsection, we investigate which kind of stable L\'evy noise has a significant  impact on the transcriptional activities, via the maximal likely concentration state value at $t = 50$  (see Fig \ref{Fig10}(a)) on the maximal likely evolution trajectories and the tipping time  (see Fig \ref{Fig10}(b)) for these trajectories.

 The red region in Fig \ref{Fig10}(a) presents the valid transcription region corresponding to the gene regulation system with the noise within the   parameter plane $(\beta, \alpha)$, as it shows the high concentration values (indicating transcription).   Note that  the symmetric ($\beta=0$)  L\'evy noise induces transition to transcription,  but asymmetric  ($\beta \neq 0$)  L\'evy noise with certain parameters do not trigger the switch to transcription.

 The dark blue region indicates the situation with no transcription.

 Likewise, as seen in Fig \ref{Fig10}(b) about tipping time in the noise case in the parameter plane $(\beta, \alpha)$, the red region illustrates that no transcription has occurred by the time $t = 50$ as there is no tipping from the low concentration to the high concentration state (i.e., the maximal likely trajectory does not pass through the saddle state $x_u$).

 In addition, the critical line $\alpha = 1$  helps form a part of the boundary between the transcription and no-transcription regions in the parameter plane. With this divided parameter plane, we can select combined parameters  $\alpha$ and $\beta$, in order to achieve transcription within an appropriate time scale.

\section{Conclusion}

In this work, we have investigated  the transcription   factor activator's concentration evolution  in a prototypical gene regulation model, focusing on the effects of Gaussian Brownian noise and non-Gaussian L\'evy noise in the synthesis reaction rate. We examine the most probable transition pathways under Gaussian noise and maximal likely trajectories under non-Gaussian noise, i.e.,  we visualize the trajectories  from low concentration to high concentration. We also compute the tipping time from the low concentration state to (or arriving near) the high concentration state. The most probable transition pathways are computed by numerically solving  a  two-point boundary value problem. The maximal likely   trajectories are calculated via numerically solving the nonlocal Fokker-Planck equation for the stochastic gene regulation model \eqref{eq:2.1}.

For a  gene regulation system under Gaussian noise,  we examine  the most probable transition pathways $z_m(t)$  from the low concentration metastable state $x_-$  to the high concentration metastable state  $x_+$, by minimizing the Onsager-Machlup action functional.
 For this same gene regulation system under non-Gaussian noise,  the Onsager-Machlup least action principle is not yet available and we thus compute the maximal likely evolution trajectory $x_m(t)$ starting from the low concentration metastable state  $x_-$. Both enable us to visualize the progress of  the transcription   factor activator's concentration evolution as time goes on (i.e., observe whether the system enters the transcription regime).


We have indeed observed that the most probable transition pathway exists under Gaussian noise for certain evolution time scale and system parameters.
Furthermore, we have characterized the concentration evolution with varying noise parameters: non-Gaussianity index $\alpha$, skewness index $\beta$ and noise intensity $\epsilon$.  Therefore, we can predict the concentration level (or an appropriate transcription status) at a given future time,  depending on the specific noise parameters in divided regions in the parameter plane (see Fig  \ref{Fig10}). We have
also noticed some peculiar or counter-intuitive phenomena. For example,  a  smaller noise intensity  may trigger the transcription process, while a larger noise intensity  can not,   in this gene  system with the same asymmetric L\'evy noise  (see Fig \ref{Fig7}(b) ). This phenomenon does not  occur in the case of symmetric L\'evy noise. Moreover,    the symmetric ($\beta=0$)  L\'evy noise induces transition to transcription for all  non-Gaussianity index $\alpha$,  but asymmetric  ($\beta \neq 0$)  L\'evy noise with certain non-Gaussianity index $\alpha$  do not trigger the switch to transcription.

These findings may provide helpful insights for further experimental research, in order to achieve or to avoid  specific  gene transcriptions.

\section*{Acknowledgments}
We would like to thank Yayun Zheng, Jintao Wang, Xu Sun, Ziying He and Rui Cai for helpful discussions. This work was partly supported by the National Science Foundation Grant No. 1620449, the National Natural Science Foundation of China Grant Nos. 11531006 and 11771449, and the Fundamental Research Funds for the Central Universities, HUST, No. 0118011075.

 \section*{Appendix}
 \label{SI}

\renewcommand{\theequation}{A.\arabic{equation}}
\renewcommand{\thefigure}{A.\arabic{figure}}
\setcounter{equation}{0}
\setcounter{figure}{0}

We recall the definition of a scalar stable L\'evy motion $L_t^{\alpha, \beta}$
and the nonlocal Fokker-Planck equation for the probability density evolution of the solution to the stochastic system (\ref{eq:2.1}).

\paragraph*{A1 Asymmetric stable L\'evy motion $L_t^{\alpha, \beta}$ }
\label{S1_Appendix}

 A scalar stable L\'evy motion $L_t^{\alpha, \beta}$ is a stochastic process with the following properties \cite{Applebaum2009,Duan2015,Sato1999,Samorodnitsky1994}: \\
(i)  $L_0^{\alpha, \beta} = 0$, almost surely (a.s.);\\
(ii) $L_t^{\alpha, \beta}$ has independent increments;\\
(iii)$L_t^{\alpha, \beta}$ has stationary increments: $L_t^{\alpha, \beta}-L_s^{\alpha,\beta} \sim S_\alpha((t-s)^\frac{1}{\alpha}, \beta, 0)$, for all $s$ and $ t$ with   $0 \leq s \leq t $;\\
(iv) $L_t^{\alpha, \beta}$ has stochastically continuous sample paths, i.e., for every $s > 0 $, $L_t^{\alpha, \beta} \rightarrow L_s^{\alpha, \beta}$ in probability, as $t\rightarrow s$.

\begin{figure}[!htp]
\begin{minipage}{0.48\linewidth}
\leftline{(a)}
\centerline{\includegraphics[height = 6cm, width = 6.5cm]{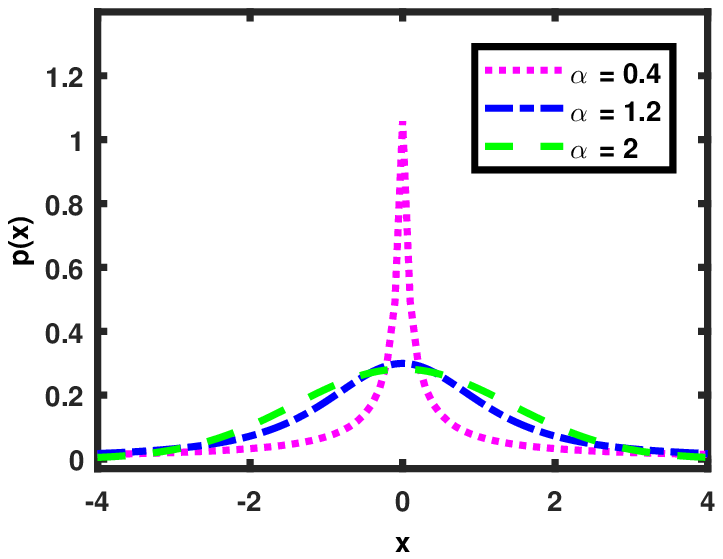}}
\end{minipage}
\hfill
\begin{minipage}{0.48\linewidth}
\leftline{(b)}
\centerline{\includegraphics[height = 6cm, width = 6.5cm]{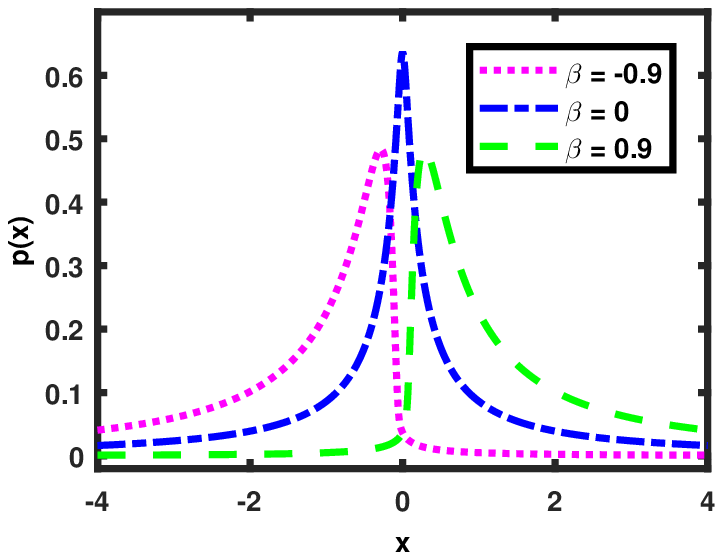}}
\end{minipage}
\caption{\textbf{(Online color) The probability density function $p(x)$ of $L_t^{\alpha, \beta}$ at time $t=1$.}
(a) Non-Gaussianity indexes $\alpha = 0.4,~1.2,~2$ for $\beta = 0$. (b) Skewness indexes $\beta = -0.9,~0,~0.9$ for $\alpha = 0.5$.}
\label{A1_Fig}
\end{figure}

Here $S_{\alpha}(\sigma,\beta,\mu)$ is the so-called stable distribution  \cite{Duan2015,Samorodnitsky1994} and  is determined by four indexes: non-Gaussianity index $\alpha   (0 < \alpha < 2)$, skewness index $\beta  (-1\leq \beta \leq 1)$, shift index $\mu  (-\infty < \mu < +\infty)$ and scale index $\sigma  (\sigma > 0)$.
A stable L\'evy motion has  jumps  and its probability density function has heavy-tails (i.e., the tails  decrease for large spatial variable like a power function \cite{Duan2015,Sato1999,Samorodnitsky1994}). The non-Gaussianity index $\alpha$ decides the thickness of the tail,  as shown in Fig \ref{A1_Fig}(a). As seen  in Fig \ref{A1_Fig}(b), the skewness index $\beta$ measures the asymmetry (i.e., non-symmetry)  of the probability density function. The distribution is right-skewed if $\beta >0$, left-skewed if $\beta < 0$, and symmetric for $\beta = 0$ \cite{Duan2015,Sato1999,Samorodnitsky1994}.

Especially, Brownian motion $B_t$  (corresponding to $\alpha=2, \beta=0$)  has  light tails (i.e., the tails   decrease   exponentially fast).

\begin{figure}[!htp]
\begin{minipage}{0.48\linewidth}
\leftline{(a)}
\centerline{\includegraphics[height = 6cm, width = 6.5cm]{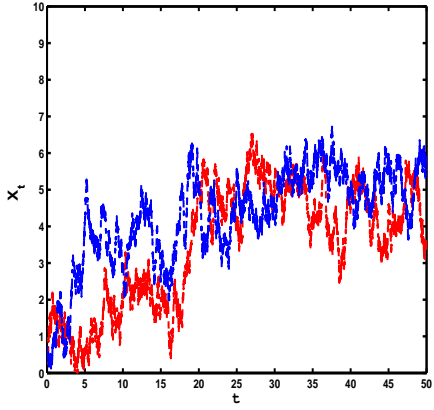}}
\end{minipage}
\hfill
\begin{minipage}{0.48\linewidth}
\leftline{(b)}
\centerline{\includegraphics[height = 6cm, width = 6.5cm]{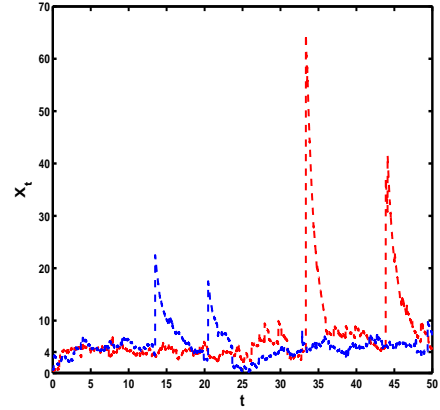}}
\end{minipage}
\caption{\textbf{(Online color) Two stochastic TF-A concentration sample paths $X_t$  starting at  $ X_0 = x_{-} \approx 0.62685$ with noise intensity $\epsilon =1$.}(a) Stochastic TF-A concentration sample paths $X_t$ with  (Gaussian) Brownian motion.  (b) Stochastic TF-A concentration sample paths $X_t$ with (non-Gaussian) L\'evy motion at $\alpha = 1.2, \beta = 0$.}
\label{Fig3}
\end{figure}

A path for $L_t^{\alpha, \beta}$, although stochastically continuous, has occasional (up to countable) jumps for almost all samples (i.e.,  realizations), while almost all paths of Brownian motion $B_t$ are continuous in time. Figs \ref{Fig3}(a) and \ref{Fig3}(b) show two sample paths of $X_t$  for the stochastic gene regulation system (\ref{eq:2.01}) and \eqref{eq:2.1}, starting at low concentration state $x_{-} \approx 0.62685$.
Unlike the phase portraits for deterministic dynamical systems \cite{GH, Wiggins}, these sample paths (`orbit' or `trajectories') mingled together and can not provide much information. Most probable phase portraits  \cite{Zhuan, Duan2015, Wang2018} for the stochastic gene regulation system (\ref{eq:2.1}) is a powerful tool to understand the stochastic system.

\paragraph*{A2  Nonlocal Fokker-Planck equation  }
\label{S2_FPE}

For the stochastic gene regulation system (\ref{eq:2.1}), let us recall the Fokker-Planck equation  for the probability density function $p(x, t) \triangleq  p(x,\! t; \; x_0, \!0)$ of its solution $X_t$, with initial condition $X_0=x_0$. The generator $A$ for the solution process $X_t$ is (\cite{Smolen1998,Applebaum2009,Duan2015})
\begin{equation}\begin{array}{l}
Ap(x,t)=(f(x)+ \epsilon M_{\alpha, \beta})\partial_x{p(x,t)}\\[2ex]+\displaystyle \epsilon \int_{\mathbb{R}^{1}\backslash \{0\}}\left[p(x + y,t) - p(x,t) - I_{\{|y|<1\}}(y)y\partial_x{p(x,t)}\right] \nu_{\alpha, \beta}(dy),
\end{array}\end{equation}
where $I$ is the indicator function and
$$
M_{\alpha, \beta} =
\left \{
  \begin{array}{ll}
    \frac{C_1-C_2 }{1-\alpha}, & \hbox{$\alpha \neq 1$,} \\
    \left(\int_{1}^{\infty}{\frac{\sin(x)}{x^2}}dx+\int_{0}^{1}{\frac{\sin(x)-x}{x^2}}dx\right)\left(C_2-C_1\right), & \hbox{$\alpha = 1$.}
  \end{array}
\right.
$$
Then the nonlocal Fokker-Planck equation for stochastic gene system \eqref{eq:2.1} is
\begin{equation}\label{eq:2.2}
  \frac{\partial}{\partial t}{p(x,t)} = A^{\ast}p(x,t),~~~~~p(x, 0) = \delta(x - x_0),
\end{equation}
where $A^{\ast}$ is the adjoint operator of $A$ and $\delta$ is the Dirac function. The adjoint operator $A^{\ast}$ can be further written as
\begin{equation}\begin{array}{l}
A^{\ast}p(x,t)= -{\partial}_x \left((f(x)+\epsilon M_{\alpha, \beta})p(x,t)\right)\\[2ex]
\displaystyle + \epsilon \int_{\mathbb{R}^{1}\backslash \{0\}}\left[p(x+ y,t) - p(x,t) - I_{\{|y|<1\}}(y)  y \partial _x p(x,t)\right]\nu_{\alpha, -\beta}(dy),
\end{array}\end{equation}
where
 $\nu_{\alpha, \beta}(dy)=\frac{C_1 I_ {\{0<y<+\infty\}}(y)+C_2 I_{\{-\infty<y<0\}}(y)}{\mid y\mid ^{1+\alpha}}dy,$
$ C_1 =\frac{H_\alpha (1+\beta)}{2}$, $C_2= \frac{H_\alpha (1-\beta)}{2}.$
Here,
 $$
H_{\alpha} =
\left \{
  \begin{array}{ll}
   \frac{\alpha(1-\alpha)}{\Gamma(2-\alpha)\cos(\frac{\pi \alpha}{2})},~~~&\alpha \neq 1, \\
   2/\pi, ~~~&\alpha=1.
  \end{array}
\right.
$$
For a symmetric stable L\'evy motion ($\beta=0$), the jump measure  is $\nu_{\alpha, 0}(dy) = \frac{H_{\alpha}}{2|y|^{1+\alpha}}dy$.

This nonlocal equation can be numerically solved by a similar finite difference method as in
\cite{Gao2016}.



\end{document}